\documentclass[aps,prc,groupedaddress,twocolumn,showpacs,floatfix,nofootinbib]{revtex4}
\usepackage{graphicx}
\usepackage[dvips]{color}
\usepackage{colordvi}

\def\beq{\begin{equation}}
\def\eeq{\end{equation}}
\def\ba{\begin{eqnarray*}}
\def\ea{\end{eqnarray*}}
\newcommand{\bea}{\begin{eqnarray}}
\newcommand{\eea}{\end{eqnarray}}

\begin{document}

\title{Proton charge radius from electron scattering}

\author{Ingo Sick}
\affiliation{Dept.~f\"{u}r Physik, Universit\"{a}t Basel,
CH4056 Basel, Switzerland}

\date{\today}
\vspace*{5mm}

\begin{abstract}
The rms-radius $R$ of the proton charge distribution is a fundamental quantity
needed for precision physics. This radius, traditionally determined from
elastic electron-proton scattering via the slope of the Sachs form factor
$G_e(q^2)$ extrapolated to momentum transfer $q^2$=0, shows a large scatter.
%
    We discuss  the approaches used to analyze the e-p data, partly redo
these analyses in order to identify the sources of the discrepancies, and
explore alternative parameterizations.    The 
problem lies in the model dependence of the  parameterized $G(q)$ needed for the 
extrapolation. 
This shape of $G(q<q_{min})$ is closely related to the shape of the charge
density $\rho (r)$ at large radii $r$, a quantity which is  ignored in 
most analyses. When using our {\em physics} knowledge about this large-$r$
density together with the information contained in the high-$q$ data, the model 
dependence of the extrapolation is reduced and
different parameterizations    of the pre-2010 data    yield a consistent 
value for $R = 0.887 \pm 0.012fm$. 
This value disagrees with the more precise value  $0.8409
\pm 0.0004 fm$ determined  from the Lamb shift in muonic hydrogen.
\end{abstract}
\pacs{14.20Dh,21.10.Ft,25.30.Bf}

\email{ingo.sick@unibas.ch}
\maketitle

\section{Introduction}

 The interest in  the root-mean-square (rms) radius $R$ of the proton charge 
 distribution is twofold: First, $R$ is  an integral quantity that characterizes
  the size of an elementary particle, the proton. Second, an accurate value for
$R$ is required in order to precisely calculate transition energies in the
hydrogen atom,  needed in connection with the definition of fundamental
constants, the Rydberg constant in particular \cite{Mohr10}, and precision tests
of QED. Traditionally, $R$ has been obtained from data on elastic electron
scattering on the proton. More recently, $R$ has been extracted from the Lamb
shift measured for muonic hydrogen. The data on transition energies in
electronic hydrogen have become so precise that $R$ can also be obtained from
measurements in electronic hydrogen, combined with fundamental constants known
from other sources.

The determination of $R$  has attracted much attention during the last years.
The value of $R$ from electron scattering --- a recent compilation listed 
0.879$\pm$0.009$fm$ \cite{Arrington15} ---   disagrees with the    more
precise value from muonic hydrogen, 0.8409$\pm$0.0004$fm$   
\cite{Pohl10a,Pohl16,Peset15}; the comparison to the radius from electronic
hydrogen \cite{Beyer17,Fleurbaey17} is not yet conclusive.     This so-called
''proton radius puzzle'' has generated an extensive discussion ranging from a
reevaluation of the uncertainties of $R$ from the determination via electron
scattering to understanding the difference in terms of new physics.  In this
paper, we will restrict the attention to electron scattering.

While the situation concerning the data base on cross sections for
electron-proton scattering is rather stable, the extraction of a radius from the
data still seems to be in a state of flux. Different types of analyses are being
carried out, and   yield contradictory results spanning the range 0.84 to
0.92$fm$, with typical error bars around 0.015$fm$. This is indicative of
a pronounced model dependence.  

In the following, we will summarize the situation on the determination of $R$ via
electron-proton scattering and provide a critical analysis of the extractions of
$R$ described in the literature; in some cases we repeat analogous
determinations to better understand the origins of discrepant results.

\section{Electron scattering}

  The electric and magnetic Sachs form factors $G_e(q)$ and $G_m(q)$ are 
determined from the cross
sections measured at given value of the momentum transfer $q$ and  scattering 
angle $\theta$ via 
 \bea
 \frac{d\sigma}{d\Omega} = \sigma_{M}~ f_r \left[(G_{e}^2 + \tau G_m^2 
)/(1+\tau) + 2 \tau ~G_{m}^2~ tg^2(\theta/2)\right] 
\label{sig} 
\eea
with $\tau$=$q^2/4m^2$, $m$ being the proton mass, $f_r$ being a
kinematical 
factor close to 1 accounting for the recoil of the proton, and $\sigma_{M}$
being the Mott cross section for scattering from a point-charge. The momentum
transfer is given by 
\bea
q^2 = 4~E~E'~sin^2(\theta/2),
\eea
$E$ and $E'$ being the incident and scattered electron energies, respectively.

The cross section depends on {\rm two} quantities, $G_e$ and  $G_m$; they can be
determined individually  via the so-called Rosenbluth separation if cross
sections at a given  $q$  are available over a large range of $\theta$. This
separation is difficult at low $q$ where $G_e$ dominates, and at  large $q$
where $G_m$ dominates.  This produces large uncertainties for the sub-dominant
form factor. During the last decade, it also became feasible to measure the
polarization transfer in scattering of longitudinally polarized electrons; the
ratio of transverse and longitudinal polarization of the recoil proton yields
the ratio $G_e / G_m$, which particularly at large $q$ helps to more accurately
determine $G_e$.

Equation (\ref{sig}) is valid in the one-photon exchange
limit (PWIA). Two-photon exchange comes from two sources: Coulomb distortion
(exchange of an additional soft photon) is important mainly at low electron
energies and changes the cross section by a few percent. Inclusion of the
correction leads to an increase of $R$ by $\sim 0.01fm$, as calculated
 in \cite{Sick98,Rosenfelder00}. The exchange of a second
{\em hard} photon is mainly important at very large $q$,  and was calculated
by {\em e.g.} Blunden {\em et
al.}\cite{Blunden05}. The main effect of the latter is to remove the discrepancy
between values of $G_e (q)$ at very large $q$ resulting from determinations via
Rosenbluth separation and polarization transfer, respectively. For a recent
review see \cite{Afanasev17},    for experiments checking upon the two photon
exchange see \cite{Rimal17,Rachek15,Henderson17}.      

As the two-photon corrections to the cross section are reasonably small, the
standard procedure is to remove the calculated 2-photon contribution from the
cross section and then analyze the data in terms of the PWIA-expression, Equation
(\ref{sig}).

Traditionally, the form factors $G_e$ and $G_m$ were determined by analyzing
cross sections and analyzing powers at given $q$ and variable $\theta$ from
individual experiments.  A better approach, used most often today, does not
depend on cross sections measured at exactly the same $q$'s and yields more
accurate form factors. The entire set of world cross section and polarization
transfer data is fit with parameterized expressions for the two form factors
\cite{Sick03c}. The fit then yields values for  $G_e$ and $G_m$, by error
propagation one can obtain realistic values for the uncertainties $\delta G_e$
and $\delta G_m$. 

\section{Charge radius and density \label{randro}}
The topic of  this review is the charge-rms radius $R$ defined
in terms of the charge density $\rho (r)$ via 
\bea
R^2 \equiv \int_0^\infty \rho (r) ~r^4 ~4 \pi~ dr , \label{geofq}
\eea
with $\rho (r)$ normalized to 1. 

In the non-relativistic limit, with velocity of the recoil proton $v \! \ll \! c$, the
charge density is related to the electric form factor via 
\bea
G_e(q)= \frac{4 \pi}{q} \int_0^\infty \rho (r)~ sin(qr)~r~dr,
\label{gofq}
\eea
an equation that can be inverted to read 
\bea
\rho (r) = \frac{1}{2 \pi^2 r} \int_0^\infty G_e(q)~sin(qr)~ q~dq.
\label{roofr} 
\eea

\noindent
This equation normally is not exploited directly, for two reasons.

 First, extension of the integral to $q$=$\infty$ is not feasible, as the data
stop at typically  {$q_{max} \sim 12 fm^{-1}$}. As a consequence, one 
postulates a model for $\rho (r)$ or $G_e(q)$, the parameters of which are  fit
to the data on $G_e(q)$. Or, better, the parameters  are fit directly to the 
cross section+polarization transfer data. This is the standard approach used for
nuclear mass numbers A$\ge$2.


Secondly, Equations (\ref{roofr},\ref{gofq}) 
require relativistic corrections  to account for the
fact that the velocity of the recoiling proton in not $\ll \! c$. These corrections are
of two types: 

{\bf a.} The dominant correction to the non-relativistic Equation (\ref{roofr})
 results from the fact that the coordinate system relevant in the
scattering process is the Breit frame, not the nucleon rest-frame. Licht and
Pagnamenta \cite{Licht70} showed that this Lorentz contraction can be corrected 
for by changing 
$q^2$  in $G_e(q^2)$  to  $\tilde{q}^2 = q^2~(1+q^2/4m^2)$. 

{\bf b.} For  composite systems the boost operator in some theories depends in
addition on the interaction among the constituents. Different models
\cite{Licht70,Mitra77,Ji91,Kelly02} yield an additional correction multiplying
$G(q)$. These factors are all of the type $(1+q^2/4m^2)^\lambda$ with, for the
charge  form factor, $\lambda$=0 or 1.


\begin{figure}[bht]
\centering
\includegraphics[scale=0.5,clip]{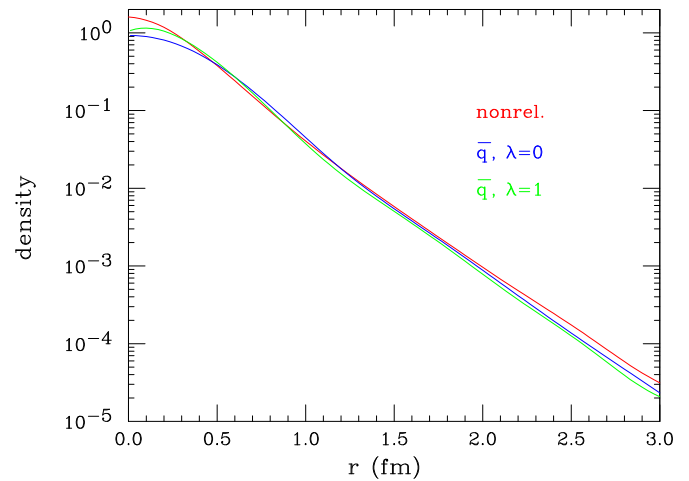}  
\caption{\label{ronrr} Densities obtained from $G_e(q)$ of \cite{Kelly04} before
(red) and
after (blue, green) application  of the relativistic corrections. }
 \end{figure}

These corrections can be  incorporated if a  quantitative density is desired. In
Figure \ref{ronrr} we show the charge density derived from a parameterized
$G_e(q)$ fit to the world data \cite{Kelly04}, before and after the  replacement
of $q$ by $\tilde{q}$ and use of the multiplicative factor. The main change
occurs for small $r$, where the density is appreciably reduced. This reduction
has a desirable effect: while densities calculated non-relativistically from
typical form factors often lead to a kink at $r$=0  --- the dipole form factor
with the corresponding exponential density is the prime example ---  the density
determined {\em after} relativistic corrections is close to flat, as it must
be.  At  $r>1fm$  the shape of the density is hardly changed,   the {\em
relative} $r$-dependence in the range 1 to 3$fm$ is changed by a factor 1.17
only. 

Figure \ref{ronrr} shows that the relativistic corrections do not  {\em
qualitatively} change the density one extracts from $G_e(q)$, and that the 
changes of the $r$-dependence  in the region of large $r$ --- particularly
relevant for the determination of $R$, see the discussion below ---  are small.
Despite relativistic corrections the density remains a valuable quantity  to
address properties of the proton and the value of $R$. We will see below that
many of the problems occurring when determining $R$ are much better understood 
--- and largely avoided --- when considering $\rho (r)$ as well; this turns out
to be true even without using explicit constraints on $\rho (r)$.

The wish to bypass the above relativistic corrections  is the reason why  the
proton rms-radius in the literature is normally hoped to be accessible via
 the slope of  {$G_e(q)$} at $q$=0 
\bea 
R^2 = -6 ~\frac{dG_e(q^2)}{dq^2}\Big|_{q=0}
\eea 
without ever considering the underlying density. Restricting the
attention to this  $q=0$ property  has caused many problems in the
determination of a precise value for $R$. The crux lies in the model dependence
 of the function needed to {\em extrapolate} from the $q$-region where 
data are available {\em and}  sensitive to $R$ to $q$=0.

\section{Data}

Cross sections for electron-proton scattering have been measured over the past
50 years, and an extensive set of data is available  
%
%
\cite{Bumiller61,Janssens66,Borkowski74,Borkowski75,Simon80,Simon81%
,Albrecht66,Bartel66,Frerejacque66,Albrecht67,Bartel67,Bartel73,Ganichot72,%
Kirk73,Murphy74,Berger71,Bartel70,Borkowski75b,Bernauer10b,Amroun89,Dudelzak63,%
Qattan05a,Christy04,Dutta03a,Niculescu00a,Rock92,Stein75,Goitein70,Bosted90,%
Bosted92,Sill86a,Price71,Litt70,Walker89,Walker94,Andivahis94,%
Drickey62,Punjabi05,Gayou01,Gayou02,Zhan11}.       
The cross sections have been measured using gaseous or liquid hydrogen targets
and electron beams with energies between $\sim$50MeV and $\sim$20GeV.  The range
of scattering angles extends from 8$^\circ$ to 180$^\circ$. The need to achieve
 small systematic uncertainties
requires accurate measurements of beam intensity (accumulated charge), target
thickness, spectrometer acceptance and detector efficiency. The most accurate
experiments have been performed at low momentum transfer, where overall
systematic uncertainties of order 1\% have been reached.

The overall normalization of the cross sections often is responsible for
the dominant systematic
uncertainty. As a consequence, many authors who analyze the data consider this
overall normalization as a free parameter. Floating the data leads to a
significantly lower $\chi^2$ for the entire {\em world} data set, but to larger
uncertainties of the values of the form factors and derived quantities.
 A safer alternative may be to keep the normalizations at the measured values,
and live with the larger $\chi^2$. In this case, the effect of the systematic
errors can be determined by changing in turn each set by the quoted systematic
errors, refitting the data and adding quadratically all the resulting changes.   

The cross sections are reasonably consistent; when floating the normalizations,
the pre-2010 {\em world} data (some 604  data points for $q_{max} < 10fm^{-1}$)
can be fit with a $\chi^2$ per degree of freedom  that is close to 1. We omit
the set of \cite{Botterill73} which shows oscillations and  yields a much 
too large $\chi^2$.

A recent experiment \cite{Bernauer10b} has tried to achieve significantly
smaller error bars by monitoring the product of charge times target thickness
with a second spectrometer placed at a fixed angle. This experiment has produced
some 1420 cross sections for $q < 5fm^{-1}$, with error bars of about 0.3\%.
 This approach came at the expense of introducing, for the 34
individual data sets 31 free normalization factors. 

The data from this experiment show significant deviations from the pre-2010 data
(see Figure 2 of \cite{Bernauer10b}).
The reasons for the discrepancy are
not entirely understood. One obvious problem of the data of \cite{Bernauer10b}
is due to the fact that the contribution of the target windows (5 to 15\%) was
not measured but simulated. This simulation included the radiative tail of the
window material, but did not include inelastic scattering from the window
nuclei. According to the one measured spectrum shown  \cite{Bernauer10a}
deviations of the measurements of $>$1\% must be expected.

This discrepancy between the pre-2010 and the Bernauer {\em et al.} data 
complicates the analysis of the {\em world} data. It leads to the fact that the
authors analyzing the data use one or the other set. Combination (see {\em e.g.}
\cite{Lee15,Sick12})  would require an increase of the  error bars of
\cite{Bernauer10b} by an amount that is difficult to gauge. Use of the  data
\cite{Bernauer10b} alone is disadvantageous because this set does not provide
(due to limitations in beam energy) data at low $q$ {\em and} large angle,
resulting in rather limited information on the magnetic form factor and radius.

Despite these difficulties, various analyses \cite{Sick12,Lee15} have shown that
for the {\em charge} rms-radius, the topic of main interest here,  the pre-2010
{\em world} and the Bernauer data basically agree.

The quantities measured in an electron scattering experiment do not directly
yield the cross sections. The measurements have to be corrected for the effects
of Bremsstrahlung (see {\em e.g.} \cite{Mo69,Maximon00}). The corresponding
theoretical corrections amount to typically 30\% and are believed to not
contribute too much to the final overall uncertainty of $\sim$1\%. 

During the last two decades, polarized electron beams have become standard at
many electron beam facilities. The polarization transfer to the recoiling proton
can be measured by placing a polarimeter in the focal plane of the spectrometer
used for detection of the recoil proton.  The measurement of the two
polarization components provides a very valuable additional observable which
helps to separate $G_e$ and $G_m$. For the polarization transfer
data 
\cite{Strauch03,Milbrath98,Pospischil01a,Crawford07,Ron07,Maclachlan06,Jones00,%
Jones06,Puckett10,Puckett12,Dieterich01b,Paolone10} the overall normalization is
less of an issue, as the observable of interest is a {\em ratio} of two
polarizations.

\section{Peculiarities and difficulties}

\subsection{Importance of $\rho (r)$ at large $r$ \label{larger}}

From Equation (\ref{gofq}) it follows that charge at radius $r_0$ generates a
Fourier component in $G_e(q)$ of type  $sin (q r_0)/(qr_0)$; for large $r_0$ it
produces  a curvature of $G_e(q)$ at low $q_0\sim\pi/(2 r_o)$. The curvature
of $G(q^2)$ --- the deviation from linearity in $q^2$ --- affects  $R$ when the
radius is   determined via extrapolation from  $q>q_{min}$, where data are
available {\em and} sensitive to  $R$, to $q$=0.  

\begin{figure}[htb]  
\includegraphics[scale=0.45,clip]{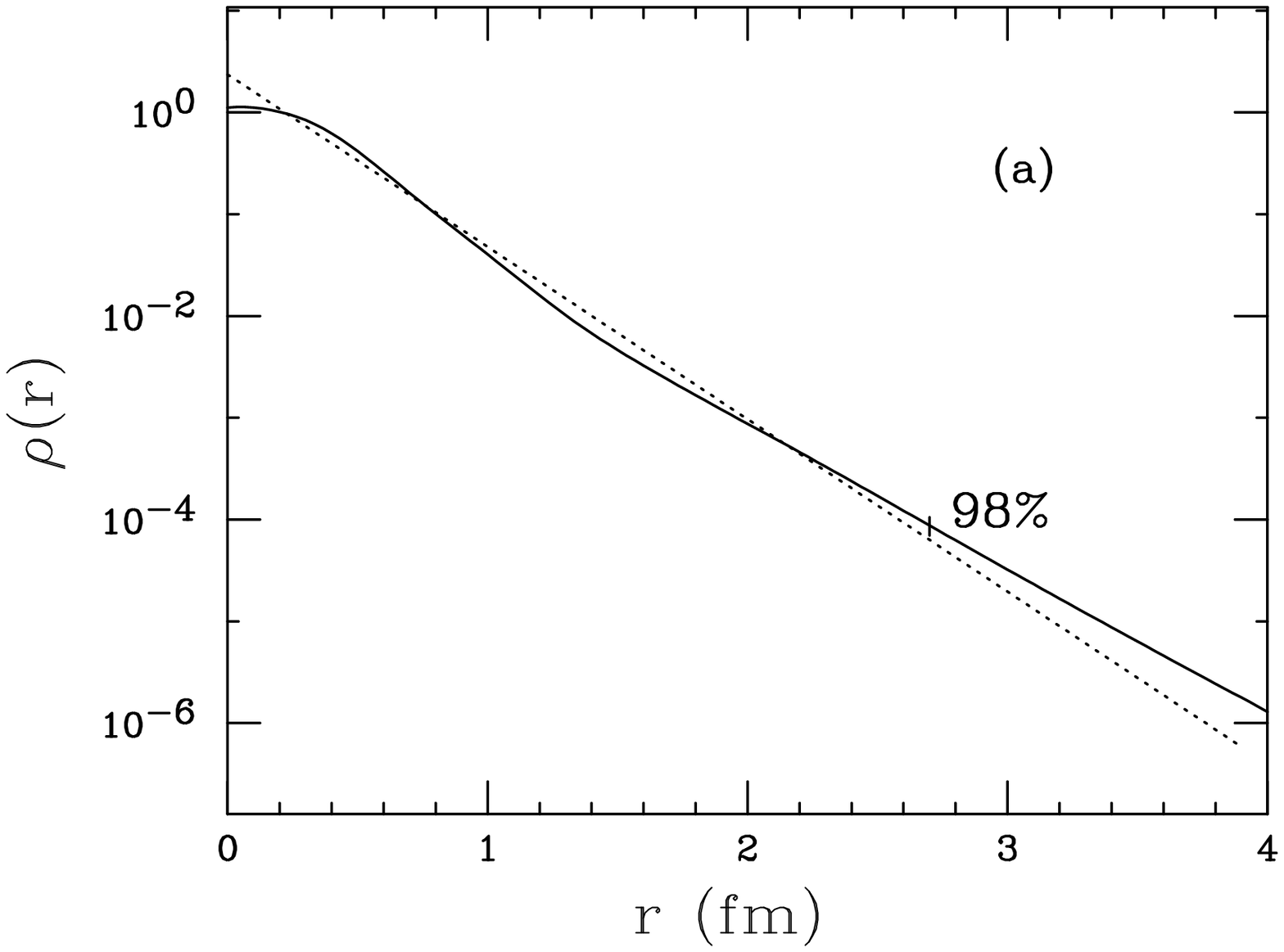}
\includegraphics[scale=0.455,clip]{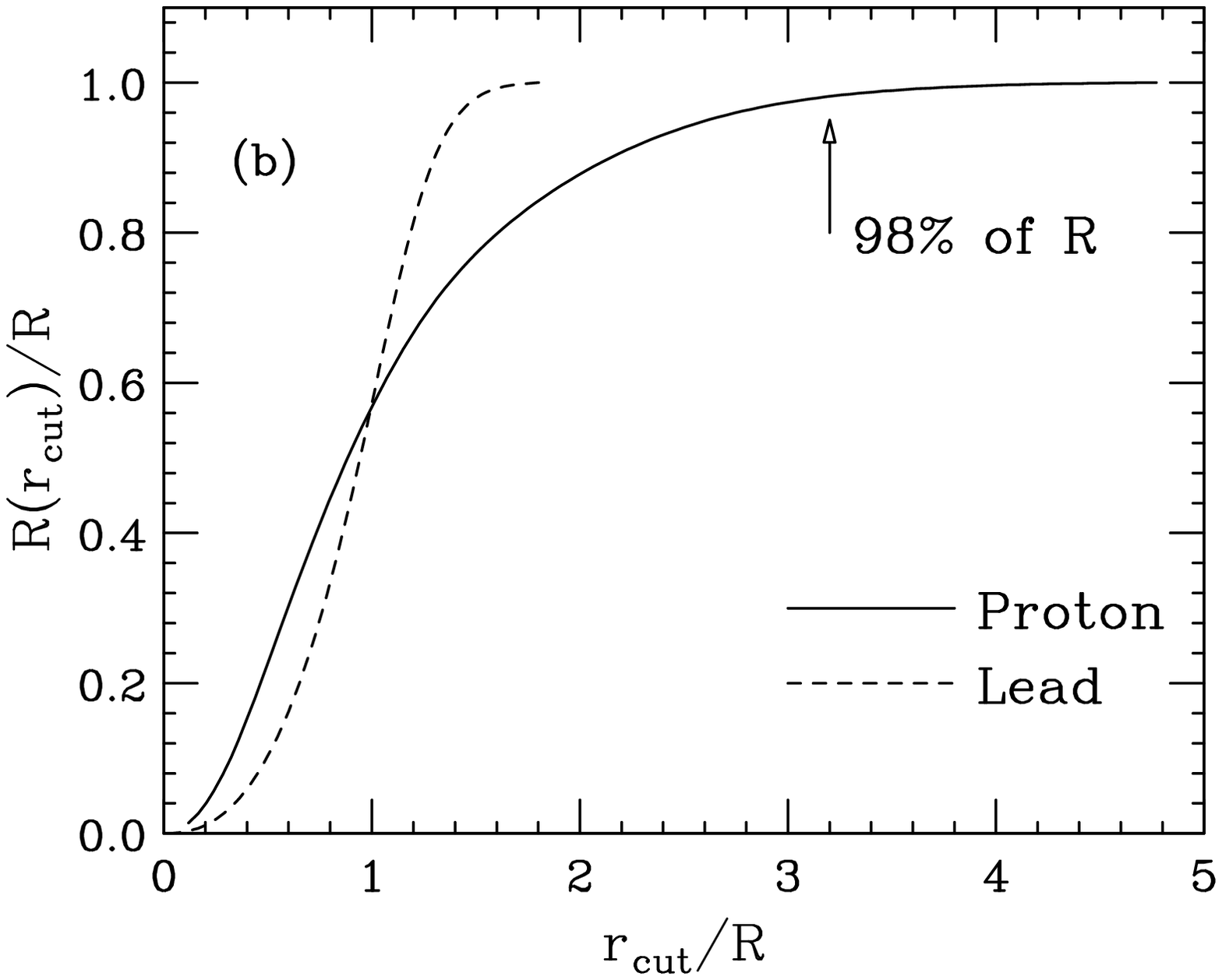}  
\caption{\label{tail} (\textbf{a}) Density (dotted = exponential, solid = 
realistic) as function of
$r$. (\textbf{b}) $R(r_{cut})/R$ as a function of $r_{cut}/R$. The result for a
heavy nucleus is shown for comparison.
}.  
\end{figure}
The charge density of the proton has a  shape that is  very different
from the typical Woods-Saxon type shapes encountered for heavier nuclei. The
 proton form factor is roughly described by the dipole shape
\bea
G_D(q) = 1/(1 + q^2 R_D^2/12)^2.
\label{dip} 
\eea
The density corresponding to this form factor has the shape of an exponential 
 \bea
\rho_D(r) \propto   e^{-\sqrt{12}~ r / R_D}.
\eea
Such a density exhibits a long tail towards large radii which contributes 
appreciably to the rms-radius.
  In Figure \ref{tail}a we show the density corresponding to a dipole form factor
(dotted) and a more realistic one (solid) resulting from the fit to the electron 
scattering data. In Figure \ref{tail}b we show the partial integral
\bea
R(r_{cut}) =\left[ \int_0^{r_{cut}} \rho (r)~ r^4 ~dr~ \left/ \int_0^\infty \rho (r)~ r^4~
dr \right]^{1/2} \right. 
\eea 
with the rms-radius given by $R$=$R(r_{cut}$=$\infty)$. To get 98\% of $R$, one has
to integrate out to 2.7$fm$, where the density has dropped to $\sim 10^{-4}$ of
the central value! 

\begin{figure}[htb] 
\centering 
\includegraphics[scale=0.5,clip]{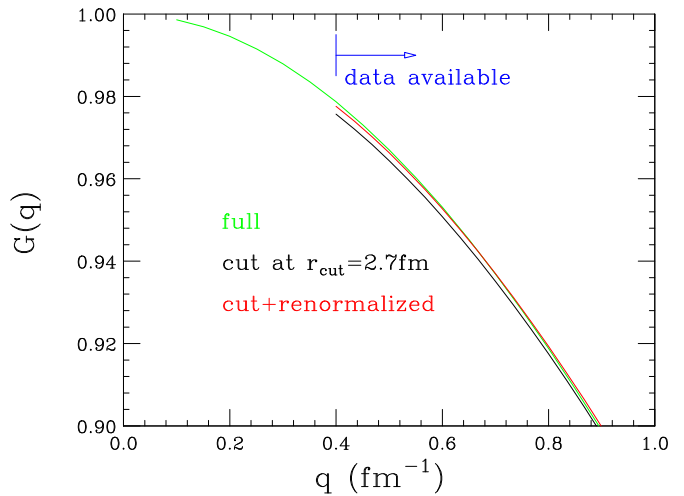}  
\caption{\label{ftail} Form factor corresponding to (\textbf{1}) full density
(green), (\textbf{2}) truncated density (black), and {(\textbf{3}) truncated}
density, renormalized to best agree with (1) (red).}
 \end{figure}

The effect of $\rho (r>2.7fm)$ upon  $G_e(q)$ at low $q$  is explored in 
Figure \ref{ftail} where we show the form factor $G_e(q)$ for 3 cases:  
\begin{enumerate}
\item   Dipole form factor (exponential density).  
\item  Form factor corresponding to exponential density truncated at 
$r_{cut}=2.7fm$. 
\item Form factor corresponding to truncated density, renormalized to agree best
with the Dipole form factor for momentum transfers above the minimum momentum
transfer of the data; this renormalization corresponds to the standard 
renormalizations of data applied in most analyses.
\end{enumerate}
The difference  between case 1 and 3 is less than 0.12\% of $G(q)$, 
which is much smaller than the uncertainties most experimentalists would claim
to be able to achieve. Due to the renormalization one would miss the
curvature of $G_e(q)$ and the contribution to $R$ from the larger-$r$ density
which, for the example chosen,  amounts to  2\%.
The same
argument could be extended to a cut at 2.4$fm$, yielding a 4\% deviation of $R$.
 We will come back to this point below. This problem can only be 
solved by constraints on the model-$G(q)$ based on the physics of the density 
at large $r$, but not by 
curve-fitting of data of realistic precision in the low-$q$ region.

\subsection{Smallness of contribution of R to $G_e(q)$ \label{small}}

When analyzing data with parameterized expressions for $\rho (r)$ or $G_e(q)$  one
needs to be aware of the potential size and location in $q$  of the main 
contribution of
$R$ to $G_e(q)$. To this end it is helpful to consider a so-called notch test. 

\begin{figure}[htb]
\centering
\includegraphics[scale=0.5,clip]{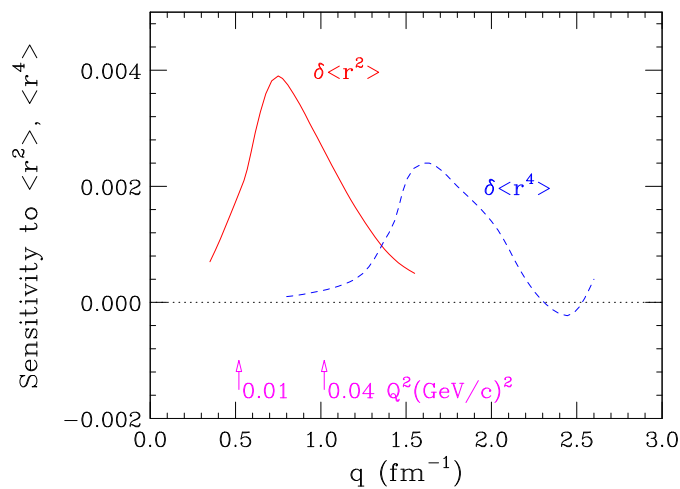}  
\caption{\label{notch} Sensitivity of  the charge rms-radius
and the 4'th moment to the e-p data at varying $q_n$.}
 \end{figure}

This test is carried out as follows: the  world data is fit with a flexible
parameterization for both charge and magnetization densities, yielding rms-radii
$R$ together with the higher moments of interest. The data in the interval 
of $\pm 0.1fm^{-1}$ around $q_n$ then is increased by 1\%. The
modified data is fit, resulting in a change of the rms-radii (and eventual
higher moments). This procedure is repeated for varying $q_n$, and the
resulting changes of the radii (and higher moments) are plotted as a function of 
$q_n$. Figure \ref{notch} shows typical results.

Figure \ref{notch}  displays the expected behavior: at very low $q$ the form
factor is
not sensitive to $R$ because of the smallness of the contribution of $R$ to 
 $G_e(q)\sim 1 -q^2~R^2/6 + ...$, at the larger $q$'s the effect of
the forth (and higher) moments dominate. The world data in the $q$-region 
$0.5 \div 1.2 fm^{-1}$ thus are mainly determining 
$R$. The region of sensitivity for
the Zemach radii \cite{Friar04,Friar05b} is quite similar. 

At the momentum transfer of maximal sensitivity to $R$, $q$$\sim$0.8$fm^{-1}$,
the contribution of $R$ to $G_e(q)$ amounts to $q^2 R^2/6 \sim 0.08$. A
determination of $R$ to 1\% then would require a knowledge of $G_e(q)$ to
$\pm$0.0016, {\em i.e.} $\pm$0.17\%. This emphasizes that a fit of the e-p data
aiming at a 1\% determination of $R$ must achieve systematic deviations from the
experimental $G_e(q)$  that are  smaller than 0.17\%. This can only be achieved
by fits that reach the smallest $\chi^2$ possible (and, in any case, smaller than
achieved by other fits). Fits that visually look good
(for examples see  \cite{Horbatsch16,Higinbotham16,Griffioen16,Lorenz14})  are
no proof of small systematic deviations from the data, simply because the plots
of data {\em vs.} fit  shown in many papers published in the past  do not {\em
by far} have the resolution that would allow  one to detect a systematic
deviation of order  0.17\%. Fits that achieve low $\chi^2$ by rescaling the
error bars of the data, as done in {\em e.g.}\cite{Griffioen16}, are not valid;
fits to exactly the same data with a $\chi^2$ lower by $\sim$300
are available.  

\subsection{Parameterizations in q-space only? \label{qonly}}
 
Due to the complications mentioned in Section \ref{randro}, most authors analyzing
the electron scattering data employ parameterizations in $q$-space only to get
the $q$=0 slope, without ever
worrying what these parameterizations would imply in $r$-space. This omission
 often leads to uncontrolled effects; in the following we discuss examples to
illustrate this point.

We have some time ago \cite{Sick14} analyzed the  data
of \cite{Bernauer10a}   for $q< 2fm^{-1}$  using a parameterized $G(q)$. Data for
$q_{max}$=2$fm^{-1}$  are usually considered sufficient  for a precise  
determination of
$R$, see  Figure \ref{notch}. The parameterization  employed was a [1/3]Pad\'e
approximant   
\bea  G(q) = \left[ 1 + a_1 q^2 \right] \left/ \left[ 1+b_1~q^2+b_2~q^4+b_3~q^6
\right] \right. .
\label{13pade}
   \eea 
With this function, the data can be fit with a $\chi^2$ which is as low as the
$\chi^2$ obtained for the same data in \cite{Bernauer10a} with a spline fit. The
Pad\'e function has none of the failures occasionally encountered in the
literature --- poles or unphysical  behavior for $q \!\rightarrow\!\infty$ --- but
it produces a charge rms-radius R of 1.48$fm$!

\begin{figure}[htb]
\centering
\includegraphics[scale=0.4,clip]{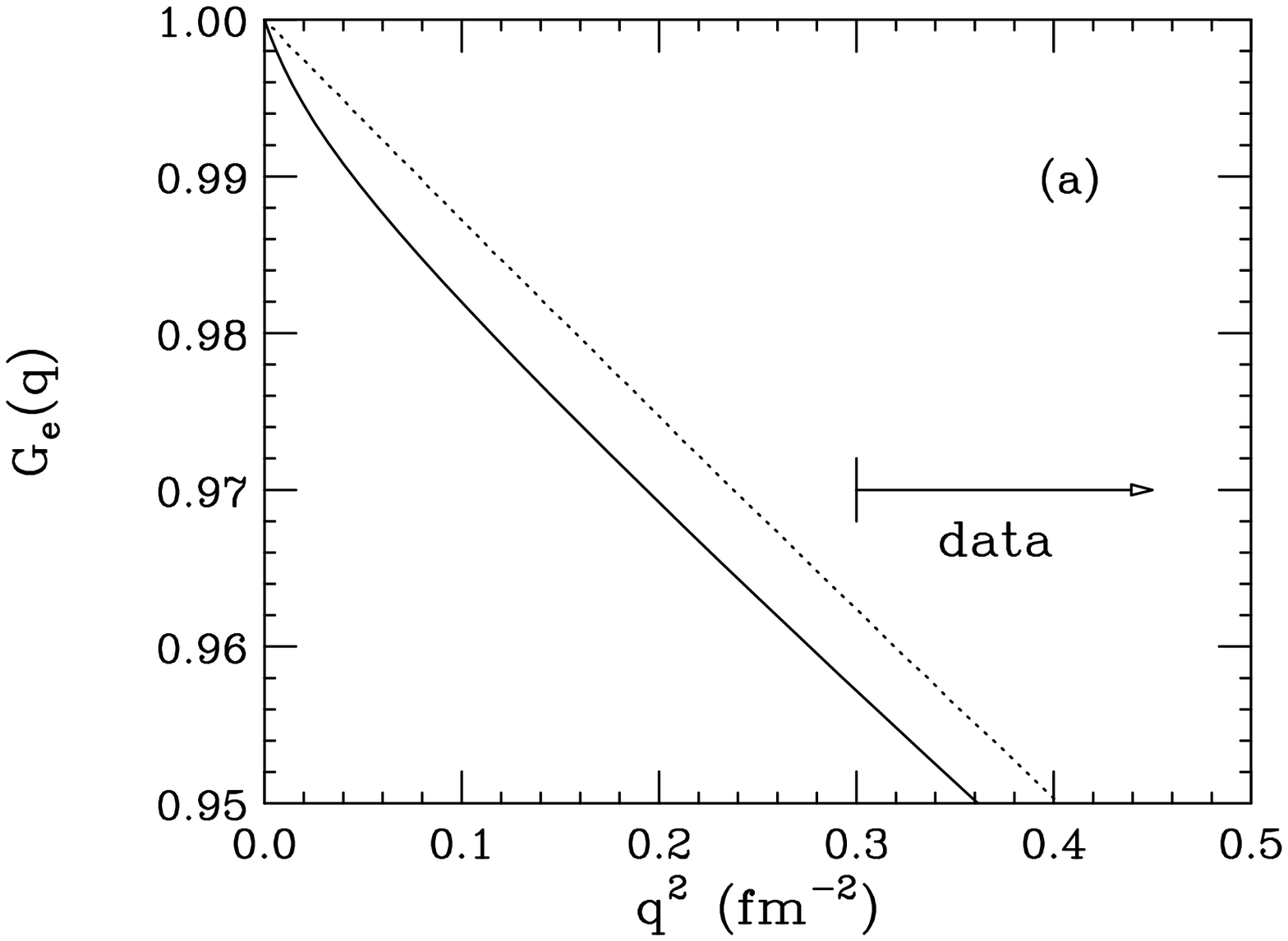}  
\includegraphics[scale=0.397,clip]{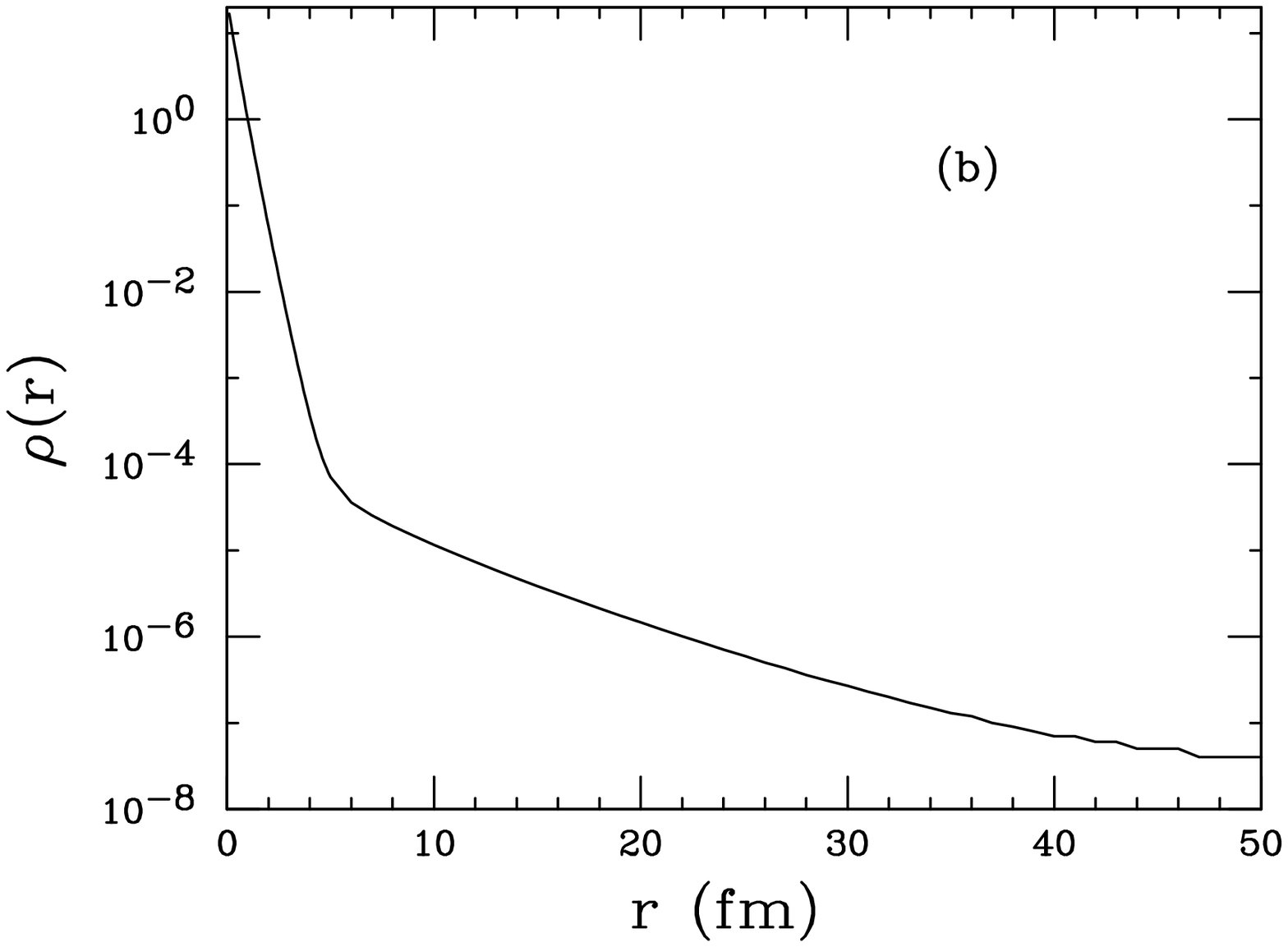}  
\caption{\label{Pade} (\textbf{a}) [1/3]Pad\'e fit (solid) together with
''standard'' fit having $R\sim0.88fm$. (\textbf{b})  Density corresponding to
Pad\'e fit.}
 \end{figure}

Figure \ref{Pade}a shows the behavior of $G_e(q)$ at very low $q$, below the range
covered by the  data. The Pad\'e fit exhibits a curvature at  $q^2 < 0.05fm^{-2}$,
leading to the large slope at $q$=0 and correspondingly large $R$.  This fit is
compared to a ''standard'' fit corresponding to $R\!\sim$0.88$fm$ (dotted). Both
fits explain perfectly the data (not shown); for $q^2>0.2fm^{-2}$ the fits differ
by an overall normalization of $\sim$1\% in $d\sigma/d\Omega$, but give the 
same $\chi^2$ as the normalization of the data is floating. 

This educational example demonstrates that it is important to examine the
density implied by the parameterized $G(q)$.  In $r$-space, the outrageous
behavior of the Pad\'e fit is immediately visible, see Figure \ref{Pade}b, and
it occurs despite the fact that the formal expression for the [1/3]Pad\'e
parameterization (Equation (\ref{13pade})) looks as acceptable as other
$q$-space parameterizations employed in the literature. The peculiar nature of
the fit results from the correlation between $a_1$ and $b_1$ which, when
assuming large values, can generate the behavior shown in Figure \ref{Pade}.

There are other examples in the literature that emphasize the importance of
considering $\rho (r)$ at the same time. Bernauer {\em et
al.}\cite{Bernauer10a}, for instance, make
an inverse polynomial fit to their data {($q_{max}\sim$5$fm^{-1}$)}. The
\begin{figure}[hbt]
\centering
\includegraphics[scale=0.4,clip]{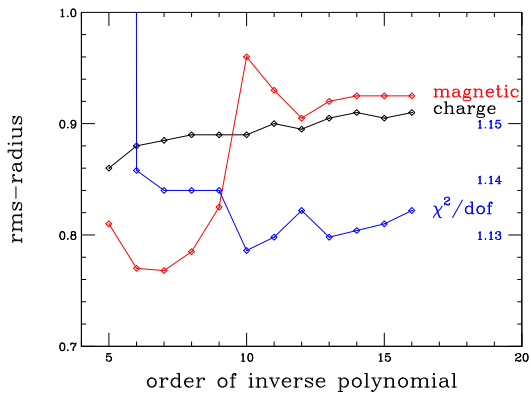}  
\caption{\label{rinvp} Charge and magnetic rms-radius from the inverse
polynomial fit, together with the $\chi^2$ per degree of freedom (right-hand
scale) \cite{Bernauer10a}}.
 \end{figure}
resulting values for $R$ as a function of the order of the polynomial are
plotted in Figure \ref{rinvp}. 
The jump of $R_m$ at order 10    (not used for the determination of $R$)   
results from a pole of $G(q)$ which happens to
occur close to  the
$q_{max}$ of the data. Such a form factor with a pole corresponds to  a density
that shows large-amplitude oscillations out to very large
values of $r$ \cite{Sick14b}, which of course affect $R$. A look at 
the density would have immediately revealed the unphysical nature of
the form factor fit.

The lesson from the above examples:  it is important to check upon the behavior
of the density implied by the chosen $G(q)$. And the  {\em most important
corollary}: it is very dangerous to employ parameterizations that do not even
correspond to a physical density, such as the large majority of published
parameterizations  which are not consistent with a large-$q$ fall-off  which is 
at least as steep as $q^{-4}$. Since the data alone are not sufficiently
accurate to fix the low-$q$ curvature, this quantity then --- in the absence of
physics constraints on the large-$r$ $\rho (r)$  --- is mainly given  by the choice of the
model for $G(q)$ used for the extrapolation;  aberrant results for $R$  such as
illustrated by the above examples then cannot be identified and excluded. One
basic problem: the parametrization can contain $sin(qr)/qr$-components (see
Equation \ref{geofq}) which
imply contributions from unphysically large radii of, say, $r>3fm$.      

\subsection{$R$ from very-low-$q$ data?}

Starting from the idea that $R$ can be determined from the $q=0$ slope of
$G(q^2)$, the form factor is often parameterized as a power series in $q^2$,
$G(q^2) = 1 -q^2 R^2/6$ plus eventual higher terms in $q^2$ (see discussion below). 
With precise data at low enough $q^2$ one could hope to determine the
$q^2 R^2 /6$ term to, say, \%-type accuracy {\em without} worrying about the
higher-order terms in $q^2$.  

\begin{figure}[htb] \centering
\includegraphics[scale=0.4,clip]{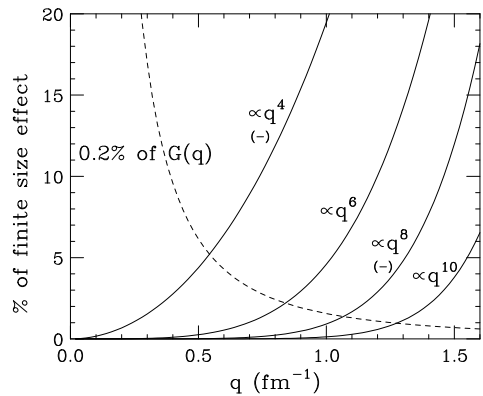}  
\caption{\label{per} Relative contribution in \% of the $\langle r^{2n} \rangle$
terms to the finite size effect $1-G(q)$, calculated using the moments of
\cite{Bernauer10a}.} \end{figure}

The problem with this approach is twofold: \\ 
\hspace*{5mm}1. Due to the
peculiar shape of the proton density the moments $\langle r^{2n} \rangle$ are
large and the $q^{2n}$-terms strongly coupled \cite{Sick03c}. This is 
illustrated in Figure \ref{per} \cite{Sick17} which shows the contributions (in
\%) of  the higher moments to the finite size effect FSE $1-G(q)$. 
In order to make the contribution of $n \ge 2$
smaller than, say,  1\% in $R$ (2\% in FSE) one has to restrict $q_{max}$ to an
extremely small value of  $\le$0.34$fm^{-1}$ (0.004$GeV^2/c^2$).  \\
\hspace*{5mm} 2. At these low $q$'s, the term of interest $q^2R^2/6$ becomes
very small  --- 0.015  at $q\sim0.34fm^{-1}$ --- but the experimental 
uncertainty of the measured quantity $G \sim 1$ remains of order 0.01. A
measurement of $R^2$ to, say, 2\% (1\% in $R$) then would require a measurement
of $G$ to 0.015$\cdot$2\%=0.03\%. Such an accuracy is  not within reach for a
very long time.  Extracting an accurate $q^2=0$ slope directly from a
measurement \cite{Gasparian11} without  dealing with the higher moments ---
without extrapolations --- is pretty hopeless.

\subsection{A counter-intuitive observation}

\begin{figure}[htb]
\centering
\includegraphics[scale=0.4,clip]{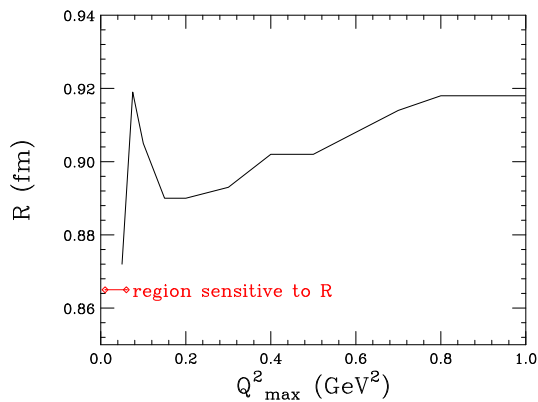}  
\caption{\label{lee} Proton rms-radius as a function of $q_{max} \leq 5 fm^{-1}$
 of the data, from
\cite{Lee15}. Indicated in red is the region of sensitivity to $R$ as 
implied by Figure \ref{notch}.}
 \end{figure}

In several of the published analyses of e-p data the dependence of the
extracted radius $R$ on the maximum momentum transfer $q_{max}$ of the data
employed  has been studied. These works have often produced an apparently
counter-intuitive result: the value of $R$  changes significantly
 with $q_{max}$. As an example we show in Figure \ref{lee} the results of Lee 
{\em et al.}\cite{Lee15}.

This dependence of $R$ on $q_{max}$ at first sight seems  incompatible with the
idea that $R$ is obtained from the slope of $G_e(q)$ at $q$=0.  Why do the 
high-$q$ data have such an impact?  It can be understood once one realizes that
the basic difficulty of the determination of the $q$=0 slope rests in the {\em
extrapolation} from $q$'s, where the form factor is sensitive to $R$ (see Figure
\ref{notch}), to $q$=0. The extracted $q$=0 slope depends on the curvature of
the parameterization at the $q$'s below $\sim$0.6$fm^{-1}$. The more we know
about the density, the better we can constrain this curvature of the 
extrapolating function. In particular, knowing more about the shape of the
density  {\em including} its large-$r$ tail  fixes better the shape of the form
factor at {\em low} $q$, as explained in Section \ref{larger}. In order to fix
best the shape of the density, the density should  explain the full set of
$G_e(q)$ data.  And it can easily be made plausible that fixing the density {\em
including} the large-$r$ tail region  requires fitting  the data up to the 
{\em largest} momentum transfers.

\begin{figure}[htb]
\centering
\includegraphics[scale=0.4,clip]{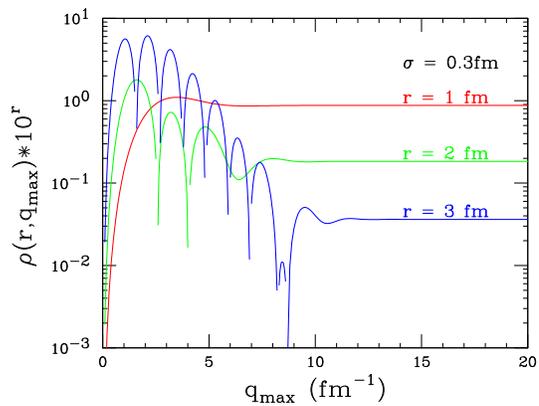}  
\caption{\label{roofq} Average density in region centered around radius $r$ as
obtained from the truncated integral over $G_e(q)$. }
 \end{figure}

 We illustrate this point with a pedagogical example. Consider the density as the
truncated Fourier transform of $G_e(q)$, defined as    \bea  \rho (r,q_{max}) =
\frac{1}{2 \pi^2 r} \int_0^{q_{max}} G_e(q)~sin(qr)~ q~dq ,  \eea  with $\rho
(r)$=$\rho (r,\infty)$.  In Figure \ref{roofq} we show $\rho (r,q_{max})$
calculated for three values of $r$  using the standard dipole form factor,
Equation (\ref{dip}). To reduce the  oscillations resulting from the sharp
cut-off of the integral at $q_{max}$, the $\rho (r,q_{max})$  is averaged over a
region around the selected $r$ with a gaussian weight of width  $\sigma$, such
that the values plotted represent some average density in the region centered at
$r$; obviously it  is only this average density which is relevant for the shape
of $G_e(q)$ at  low $q$.

Figure \ref{roofq} shows that  for  $r$=1$fm$ the (averaged) density is well
determined once the upper limit of the integral has reached about  $5fm^{-1}$.
For larger $r$, higher $q_{max}$ is needed; the density in the  $r=3fm$ region
is only reasonably determined once $q_{max}\sim$11$fm^{-1}$. Due to the
smallness of $G(q)$ at large $q$ the high-$q$ data help to determine the
small-amplitude Fourier components which are important to fix the small
densities occuring at large $r$. Figure \ref{roofq} shows why fits that explain
the data up to larger $q_{max}$ fix better the density at larger $r$, hence
yield (implicitly) a more realistic shape  of the $G_e(q)$ needed to extrapolate
to $q$=0.  We therefore in this review concentrate on the pre-2010 data which
reach the higher $q_{max}$.

To summarize: the extrapolation  of low-$q$ data to $q$=0 is most often based on
model-dependent  parameterization of $G(q)$.  {\em Much
more reliably}, the shape of $G(q)$ can be constrained by  the shape of $\rho
(r)$ at large $r$ known (implicitly) from fits to data up to the highest $q$'s,
and these fits produce the most trustworthy values of $R$. As will be shown 
in Sections \ref{inv}-\ref{bor}  fits including the large-$q$ data yield
large-$r$ tails of $\rho (r)$ that are ''well-behaved'', {\em i.e.} are close to
the one obtained with a physics constraint as described in Sections 
\ref{md}-\ref{sog}.  

\section{Parameterizations and fits}
The data on electron-proton scattering have been analyzed in the past by a
number of authors. Different procedures have been employed, a variety of
parameterizations have been used.  Below, we discuss a representative
set, from which in the end we aim to distill a reliable value for the
rms-radius.

\subsection{Types of parameterizations used \label{type}}

 1. For the interpretation of data at very low values of $q$ various traditional
expressions, depending on one or two parameters, have been used: dipole, double
dipole, gaussian, Yukawa ...  , see {\em e.g.} \cite{Bernauer14,Horbatsch16}. 
Only those  parameterizations are retained that give a $\chi^2$ close to the
minimal one found. The obvious risk of this approach: parametrizations with too
few degrees of freedom yield too large $\chi^2$ and unreliable $R$ 
\cite{Griffioen16,Higinbotham16,Horbatsch16}. 
Figure \ref{per} can be used to estimate how many independent parameters 
(moments) are needed to achieve a given accuracy of $R$ for a given $q_{max}$.

2. When fitting data up to large $q$, which requires many free parameters, a 
different approach is needed. Multi-parameter models such as the Pad\'e form
factors \cite{Arrington07,Sick03c}, polynomials or inverse polynomials of high
order \cite{Bernauer14,Kraus14,Arrington07}  or polynomials as a function of
derived quantities \cite{Hill10,Lee15,Borisyuk10} have been employed. 
Typically, the number of parameters is increased until the $\chi^2$ per degree
of freedom  reaches a plateau. Occasionally, the model dependence is estimated
by generating and fitting pseudo-data, and comparing the fit-results  to the
known input values  \cite{Sick03c,Borisyuk10,Kraus14,Bernauer14}. 

3. A somewhat more systematic approach employs an expansion of the form factors
on an orthogonal basis \cite{Kelly02,Friar73,Anni95}. This eases the
determination of the parameters, but the  selection of the appropriate cut-off
in the order of the expansion  --- mostly based on the  $\chi^2$-plateau
argument ---  is more delicate.  The use of gaussian bounds on the individual
parameters, implemented by a ''penalty''-contribution to $\chi^2$
\cite{Lee15,Hill10,Graczyk14}, is also quite efficient in limiting the
values of the  highest-order coefficients, which tend to be poorly
constrained by the data.  

4. Safer approaches  try to include known physics in the parameterization,
hereby restricting the freedom of the fit. Examples are the Sum-Of-Gaussians SOG
densities which limit the fine  structure in the density \cite{Sick74}, or
semi-phenomenological Vector Dominance Model (VDM) based fits which  employ the
analytical form of the VDM-$G(q)$ and/or constrain the large-$r$ fall-off, see 
\cite{Iachello73,Blatnik74,Gari92,Lomon01,Bijker04} and  Sections \ref{md},
\ref{lag}.

5.  The strongest --- and often too strong --- input from theory is present
in approaches such as the VDM-fits, where constraints come from the assumption
of vector-dominance and the experimentally known masses and couplings  of the
vector mesons (see Section \ref{vdm}).

\subsection{Polynomials in $q$ \label{poly}}

Given that the rms-radius most often is thought of as the slope of the form
factor at $q$=0, it is popular to employ for the parameterization of the
Sachs form factors $G_e$ and $G_m$ 
\bea
G(q) = 1 +a_2 q^2+ a_4 q^4 +a_6 q^6 +..... , \label{polynomial}
\eea
with $a_2$=--$R^2$/6 and, non-relativistically, $a_4$=$\langle r^4 \rangle/120$  
and $a_6$=--$\langle r^6 \rangle/5040$ given by the higher moments of the density
distribution.

The above expression is often taken as the Taylor expansion of $G(q)$ around 
$q$=0, in which case the question of the convergence radius ($q^2$=4$m_\pi^2$ in
the VDM) can come up. Alternatively, one can take this parameterization  simply
as a polynomial fit in the selected $q$-range, in which case the
parameterization is not subject to  this concern. For a discussion see
\cite{Bernauer16a}.

We have pointed out  long time ago \cite{Sick03c} that a polynomial fit
is not suitable; the rms-radius and higher moments are very dependent on the
cut-off $q_{max}$ and the number of terms employed. This is essentially a
consequence of the fact that for an exponential-type density the higher moments
increase rapidly with order.  As a result, the convergence of $G(q)$ with
order is poor: the contributions of the higher terms grow and are alternating in
sign (see also Figure \ref{per}).

\begin{figure}[bht]
\centering
\includegraphics[scale=1.0,clip]{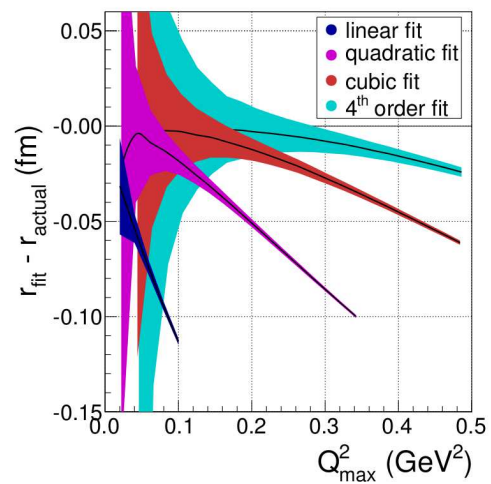}  
\caption{\label{kraus} Difference between input and fit value for the rms-radius
as a function of the cutoff in $q^2$ \cite{Kraus14}.}
 \end{figure}

Kraus {\em et al.}\cite{Kraus14} have demonstrated the inadequacy of the
polynomial fits in a quantitative manner. They generated pseudo-data, in a
realistic $q$-range with realistic error bars, using different parameterized
form factors with known $R_{actual}$.  These data were fit using the above 
polynomial 
expression, and the resulting $R_{fit}$ compared to $R_{actual}$. Their result is
shown in Figure \ref{kraus} as a function of the upper limit $q_{max}$ 
 and for different orders of the polynomial. For a cutoff at
low $q_{max}$, the error bars on $R$ (shaded areas) are large, for higher cutoffs the
error bars get smaller, but the resulting values of $R_{fit}$ are systematically
below the input value. 
Polynomial fits simply give wrong results, such as the radii of
\cite{Griffioen16,Higinbotham16} in the 0.84$fm$
region, see \cite{Sick17}.

The behavior shown in Figure \ref{kraus} can  be understood qualitatively. Consider
{\em e.g.} a linear fit of  the very-low $q$ data with $G_e(q)$=1$-q^2 R^2/6$
\cite{Griffioen16,Higinbotham16}, with all
higher moments set (implicitly) to zero. The
most important neglect then is due to the lowest higher moment  $\langle r^4 \rangle$.  A
charge distribution with $\langle r^2 \rangle$ finite and 
$\langle r^4 \rangle$=0
 must have a pronounced negative tail at large $r$, needed to
reduce $\langle r^4 \rangle$ to 0. This negative tail of course also affects
$R^2$, and leads to the behavior exemplified by Figure \ref{kraus} (dark blue curve).
 The same thing happens if the parameterization is cut off at higher order.

The above reasoning on the charge density is, of course, only a plausibility
consideration. In
reality there is no physical charge distribution that could correspond to a
polynomial-type form factor, since the Fourier transform of any polynomial 
expression diverges.

Polynomials in $q$ or $z(q)$ (see Section \ref{map})  have another generic ---
and serious --- problem. The functional dependence does not naturally embody the
fact that form factors are {\em steeply falling} with increasing $q$.  As a 
consequence  the small values of $G(q)$ at large $q$  have  to be generated by
delicate cancellations including  the low-order terms, which typically give the
largest contribution to FSE near $q_{max}$. For {\em e.g.} the parameterization
of Lee {\em et al.}\cite{Lee15} (to be discussed below)  the term $a_1 z$, which
is meant to parameterize the $q^2=0$ slope and  determine  $R$, still
contributes 70\% of the finite size effect FSE=1$-G_e(q)$ at the maximum $q$ of
5$fm^{-1}$ employed, leading to undesirable correlations affecting $R$. For
polynomials in $q^2$ this disease is even worse (400\% of FSE at 5$fm^{-1}$).  

\subsection{Inverse-polynomial type \label{inv}}

The standard polynomial-type parameterization discussed above has a number of
properties that make it unsuitable. An inverse polynomial, such as used 
for instance by Bernauer {\em et al.}\cite{Bernauer14}, avoids the problem of divergence in
the $q \!\rightarrow \! \infty$ limit. But it retains the undesirable feature of
strong correlations between the individual terms, leading to coefficients with
 alternating signs that grow with order. 
Zero's of the polynomial, leading to poles in $G(q)$ 
(in general at $q>q_{max}$) addressed already in Section \ref{qonly}, and 
the steep fall-off of inverse-polynomial $G(q>q_{max})$ then lead to
densities that show oscillations out to unphysically large radii.

A variant is the Pad\'e parameterization, which often is the ''best''
approximation of a curve by a rational function of given order   
\bea
G(q) = \left( 1 + \sum_{i=1}^I a_i ~q^{2i} \right) \left/ \left(1+\sum_{j=1}^J b_j 
~q^{2j} \right) \right. . 
\eea
If the coefficients in the denominator are constrained to $b_j>0$ and the order
$J$ to $J\!\ge\! I$+2, poles and divergences are avoided. This function has been 
used by Kelly \cite{Kelly04} and Arrington {\em et al.}\cite{Arrington07}. 
The latter   authors have performed a fit to
the pre-2010 world data ($q_{max}$=10$fm^{-1}$) including two-photon corrections. 
Their fit (not especially oriented towards a determination of $R$) has an
excellent $\chi^2$ and yields $R$=0.878$fm$. The  corresponding
density (see Figure \ref{ronrr}) is ''well behaved'', {\em i.e.} falls at large
$r$ similarly to tail densities  obtained with physics constraints (to be
discussed with Figure \ref{rovdma}).   

Special cases of  the Pad\'e parameterization, in  the form of continued
fraction CF expansions, have been employed in
\cite{Sick03c,Griffioen16,Higinbotham16,Lorenz12a}. Besides the difficulties
mentioned above, finding  the  parameters of the global best-fit was sometimes 
not successful  \cite{Bernauer16a}.  CF fits with too few parameters
\cite{Griffioen16,Lorenz12a} yield values of $\chi^2$ that are much larger than
other published fits to the same data, hence  yield no reliable $R$.  

\subsection{Polynomials in  z(q) \label{map}} 

   Dispersion relations show that    the form factor is an analytic
function of $t$=--$q^2$, with a cut beginning at the two-pion threshold 
$q^2_{c}$=--$t_{c}$=--4$m_\pi^2$. 
 Hill and Paz \cite{Hill10} exploit this by mapping $q$ onto the variable  
\bea 
z(q^2)  = \left( \sqrt{t_{c}-t}-\sqrt{t_{c}-t_0} \right) / \left( 
\sqrt{t_{c}-t}+\sqrt{t_{c}-t_0} \right) \label{zzz}
\eea  
with $t_0$ usually set to 0.  At  small momentum transfer $z$ is
proportional to $q$, while at large momentum transfer  $z \!\rightarrow\! 1$.   Paz
and Hill then  expand the form factor as a power series in $z$, 
\mbox{$G(z) = \sum a_k z^k$} . For the physical region this implies the existence of 
a small expansion parameter $z<1$.

Use of $z$ maps the cut onto the unit circle in the complex variable $z$, which
is hoped to reduce the curvature in the form factor which complicates the
extrapolation to $q$=0. The curvature of \mbox{$G(z<z_{min})$} is indeed reduced by
$\sim$30\%  \cite{Hill10}, but this  does not really remove the problem of the
unknown curvature of the true form factor. 

The advantage of the expansion in terms of $z$: the coefficients multiplying
$z^k$ are bounded. Lee {\em et al.}\cite{Lee15}, who present an extensive and
very  detailed analysis of the e-p world data using the same approach, employ a
uniform bound on $a_k/a_0$,  enforced by an additive penalty
$a_k^2/a_0^2$ in the $\chi^2$. This  has the benefit of making uncritical the
number of parameters used, contrary to the standard approaches where too many
parameters can lead to an over-fitting of fluctuations of the data and error
bars that blow up. For the $z$-expansion without such bounds 
\cite{Lorenz14}, the parameters get huge and no stable solution for $R$ is
 found (for examples see \cite{Bernauer16a}).

From a careful analysis with a multitude of fits Lee {\em et al.} deduce a
charge rms-radius {$R$=0.916$\pm$0.024$fm$}. The authors attribute the
larger-than-usual value of the radius to the use of the physics constraint
introduced by the $z$-expansion.

One obvious problem with the polynomial in terms of $z$: the form factor in the
limit $q\!\rightarrow\!\infty$ goes to a constant value, of order 1. This means
that there exists  no physical density that  corresponds to this $G(q)$. It then
is not possible to check whether the curvature of $G(q)$ at $q$'s  below 
$\sim$0.6$fm^{-1}$ would correspond to a sensible large-$r$ behavior of the 
density.

In order to investigate the origin of the large rms-radius found by  Lee {\em et
al.} we have repeated their analysis using the same $z$-variable, but optionally
assumed for $G(z)$  a power series in $z$  {\em times} a dipole in $q$. This
parameterization is similar to the one proposed by Borisyuk\cite{Borisyuk10},
see below; the polynomial in this case describes only the deviation from the
(dominant) dipole $q$-dependence, while the dipole fixes the generic problem of
polynomials discussed at the end of  Section \ref{poly}.  For this
parameterization of $G(q)$ a physical density does exist and one can check
whether the density falls  at large $r$ like densities with a tail constrained
by physics  (to be discussed with Figure \ref{rovdma}). When replacing the
polynomial in $z$ by this parameterization of $G(z)$, but otherwise following
closely the approach of Lee {\em et al.},  we find a  systematic change of the
charge rms-radius   of $\sim -0.03fm$, and a  large-$r$ tail  which is  close to
the one obtained with a physics constraint as described in Sections 
\ref{md}-\ref{sog}.  

From this study we conclude that it is not the mapping into the variable $z$
that is responsible for the large $R$ of \cite{Lee15}, but rather some
hard-to-identify peculiarity of the unphysical  $G(z(q))$. 

\subsection{Polynomial in $\xi$  times dipole \label{bor}}
In analogy with the study of Lee {\em et al.}  discussed above, Borisyuk
\cite{Borisyuk10} also maps $q$ into a new variable 
\bea
\xi = q^2/(1+q^2/\xi_0),
\eea
with $\xi_0$=0.71$GeV^2/c^2$. This $\xi$ is very similar to the $z$-variable of
Lee {\em et al.}, the main difference being that it reaches $\sim0.7$ rather 
than 1
in the limit $q\rightarrow\infty$. To parameterize $G(q)$ Borisyuk uses  
\bea
G(q) = (1 -\xi/\xi_0)^2 \sum a_k~ \xi^k
\eea
which, as compared to Lee {\em et al.}, has an additional factor 
$(1-\xi/\xi_0)^2 = 1/(1+q^2/\xi_0)^2$ corresponding to the standard dipole. 
Multiplication with
this factor ensures a physical behavior of $G(q) \sim q^{-4}$ in the large-$q$
limit. 

 Borisyuk determines the optimal parameters  by looking, for given number $n$ of
terms, at the error bars of $R$  as a function of $q_{max}$ of the data. For 
decreasing  $q_{max}$ the uncertainty $\delta R$ of $R$ grows due to the
statistical errors of the data and the smallness of the finite size effect FSE,
for increasing $q_{max}$ the uncertainty grows due to increasing systematic
error resulting from a poor fit of the data; this latter contribution is
estimated by comparing $R$ from fits of pseudo-data to the $R$ used to generate
the pseudo data. In between, the uncertainty has a minimum which is taken as
$\delta R$.   For the combination $n$=5, $q_{max}$=5$fm^{-1}$  Borisyuk finds  for
$R$ a value of $0.914 \pm 0.011fm$.

In his fits, Borisyuk uses much of the pre-2010 cross section data, but does not
include  the polarization transfer data which are very important  to reliably
separate $G_e$ and $G_m$. He also does not include some of the older, yet
precise,  data at very low $q$.

Repeating his analysis with our  pre-2010 cross section data base, the
available  polarization transfer data  and the two-photon
corrections of \cite{Blunden05} we find, for $q_{max}$=5$fm^{-1}$, $R$=0.880$fm$.  
The density for $r > 2.4fm$ falls more slowly than the MD density 
(Figure \ref{rovdma}), but approaches it  when extending  $q_{max}$ to 
$10fm^{-1}$.  

\subsection{R from Bayesian inference \label{gra}} 

   
 Graczyk and Juszczak\cite{Graczyk14}  present an analysis of the world data 
that is based on a very different philosophy. In their approach, based on 
Bayesian inference, the two form factors are assumed to come from a large 
class of possible models, each described by a functional form depending on a 
number of parameters. They use neural networks with one hidden layer and a 
variable number of neurons, as such a structure is able to approximate any 
function in principle. The weights of the neural network function are then 
determined using the Bayesian theorem and as the maximum a posteriori (MAP) 
for each model. For the likelihood they used a $\chi^2$ distribution.
In order to avoid an overfitting of the experimental 
results a prior distribution on the parameters, that is the weights of the 
individual neurons, is added. This is assumed to be of a Gaussian form with 
the width optimized as well. Such a prior can be seen as a (Tikhonov) 
regularisation of the result, prefering smooth functions. The comparison 
of the different models, given by different number of neurons and therefore 
different model complexity is done by approximating the model evidence with 
the MAP value of the likelihood together with the Occam factor to take into 
account the uncertainty of these values.
\begin{figure}[bht]
\centering
\includegraphics[scale=1.0,clip]{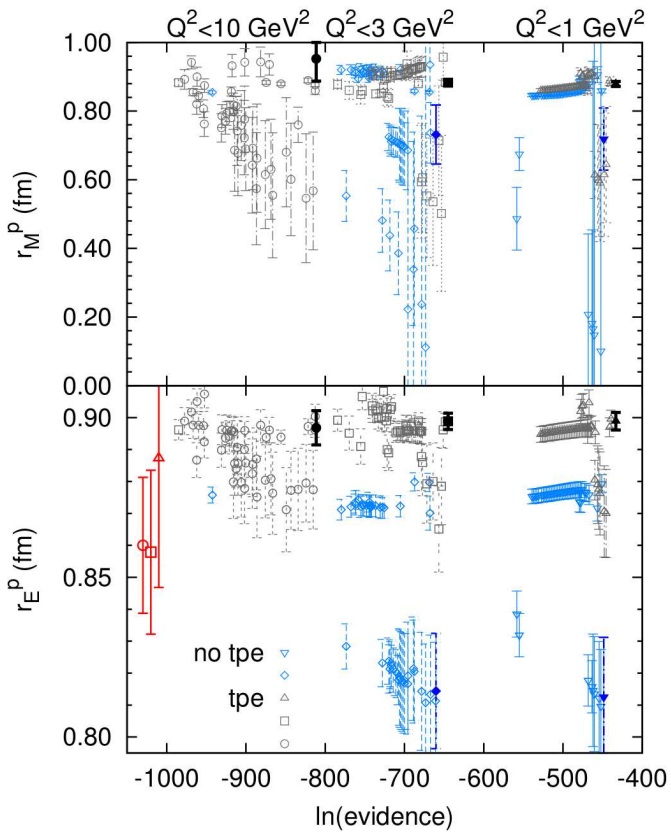}  
\caption{\label{grac} Proton radii from the Bayesian network analysis of
\cite{Graczyk14}. The points in black are corrected for the effects of two-photon
exchange. }
 \end{figure}

Graczyk and Juszczak have applied this approach to the pre-2010 world cross
section data including the polarization transfer results. The input data were
corrected for the effects of two-photon  exchange. The resulting radii, for three
different values of $q_{max}$, are plotted in Figure \ref{grac} as a function of
the evidence. The density, calculated from the coefficients given in
\cite{Graczyk10} ($q_{max}\sim$10$fm^{-1}$) , is well behaved and  out to 2.7$fm$ very close to the 
MD  and SOG densities  (to be shown in  Figure \ref{rovdma}) which
include a physics constraint at large $r$. 
From these values the authors extract the charge-rms radius which, for the three
values of $q_{max}$ of Figure \ref{grac}, amounts to 0.899(0.003), 0.899(0.003) and
0.897(0.005) $fm$, respectively. The error bars appear low given the size of the
systematic (normalization) errors of the data (see Section \ref{small}). 

\subsection{Vector Dominance Model fits \label{vdm}}

The Vector Dominance Model VDM has been used over several decades to analyze
electron scattering data \cite{Gari85,Mergell96,Hammer04,Lorenz15}, since the 
times of H\"ohler and collaborators \cite{Hoehler76,Hoehler75,Hoehler83a}. Use of
the VDM allows  to add physics to the interpretation of
the e-p data and remove much of the arbitrariness involved with purely
phenomenological parameterizations of $G(q)$ (discussed in part above).

In this model, the interaction between the virtual photon and the nucleon is
assumed to be carried by the exchange of vector mesons, with the information
contained in the spectral function that describes the strength at $q^2<0$. Most
of the parameters are obtained from the properties of the experimentally known
resonances,  some are derived via dispersion relations from other processes such
as $\pi$N scattering, some of the parameters are fit to the electron
scattering  data.  In the VDM, the resonances lead to  form factors that {\em a
priori} have the $q$-dependence of a monopole. In order to obtain the correct
high-$q$ behavior (a dropoff at least as fast as $q^{-4}$), super-convergence
constraints or an additional multiplicative dipole form factor are used. 

Here we  cannot review the many VDM calculations. Rather, we restrict our
comments to aspects that are related to the determination of the charge
 rms-radius.

Since the work of  H\"ohler {\em et al.}, all VDM analyses 
\cite{Mergell96,Adamuscin12,Lorenz12a,Lorenz14,Lorenz15} have produced fits that
have a significantly higher $\chi^2$ than phenomenological parameterizations. 
This was linked to systematic differences between data and fit at low $q$.  Note
that these deviations would be visible only when making the comparison data-fit with
the resolution required to  detect differences of a  fraction of a percent (see
Section \ref{small}) which  generally has not been done. These differences resulted in 
charge rms radii in the 0.84$fm$ neighborhood, which are low  compared to the
$\sim$0.88$fm$ from phenomenological fits. Obviously the constraint imposed by
the VDM is very strong, with the consequence that the parameterization does not
offer enough flexibility to allow perfect reproduction of the low-$q$ e-p data.

This overly strong constraint from the model is perhaps best illustrated by the
VDM analysis of Adamuscin {\em et al.}\cite{Adamuscin12}. These authors find a
radius  $R$=0.849$\pm$0.007$fm$. They obtain a very small error bar 
 despite the fact that they do  not use {\em any} e-p  cross section
data. They only employ the polarization transfer data, which measure 
the {\em ratio} $G_e/G_m$, but carry no information on $G_e$!  Obviously, the
Vector Dominance assumption  fixes $R$ all by itself.

The difficulty of the  VDM-fits to reproduce the low-$q$ data is related to
the small strength of the (isovector) spectral function right below the
$q^2 < -4 m_\pi^2$ threshold, where the ''triangle diagram'', related to the
one-pion tail,  contributes \cite{Hammer04}. The strength in this region has
been determined at the time by H\"ohler {\em et al.} via dispersion relations,
and has been used ever since. The corresponding density falls somewhat faster
than densities with better $\chi^2$, see Section \ref{sog}.

This point is also  illustrated by the recent work of Lorenz {\em et
al.}\cite{Lorenz15}. These authors compare the spectral function of the VDM to
the one obtained from a  constrained-$z$ expansion fit of the e-p data (see
Section \ref{map}). The spectral function from the latter shows pronounced 
strength near the $2 \pi$ threshold.  
Given this situation, it would be  highly desirable to redo the dispersion
analysis of H\"ohler {\em et al.}.

   Alarcon and Weiss \cite{Alarcon17} have combined the dispersion analysis
with chiral effective field theory. The latter allows to predict the higher
moments $\langle r^{2n} \rangle$ which affect $R$ when using Equation
\ref{polynomial}
 to analyse electron scattering \cite{Horbatsch17}. These higher moments are,
however, very far from the values obtained from electron scattering
\cite{Bernauer10a}.   

\subsection{VDM-motivated parameterizations \label{md}}

In the VDM the form factor is given by a sum of (an integral over) monopole
contributions. In the semi-phenomenological  analyses of data 
\cite{Iachello73,Blatnik74,Gari92,Lomon01,Bijker04}, only this analytic
structure of the basic parameterization is taken from the VDM, but most parameters 
are fit to the e-p data. These parameterizations also include a modification,
often an additional multiplicative dipole form factor, to account   for the
non-pointlike vertices and to fix the incorrect asymptotic behavior for $q
\rightarrow \infty$ of the monopole terms.  These semi-phenomenological
approaches have been highly successful in the past, and have allowed to describe
the entire set of  nucleon form factors over a large range in momentum transfer.

Motivated by this success, we have used an analogous parameterization, a sum of
Monopoles times Dipole, MD for short
\bea
G(q) = (1+q^2/M^2)^{-2} ~ \sum_{i=1}^I a_i ~(1+q^2/m_i^2)^{-1} , 
\eea
where the  $m_i$ would correspond (in the VDM) to the position of poles and
$M$ is large  compared to the $m_i$'s in order  not to affect 
 the low-$q$ behavior suggested by the VDM. The amplitudes $a_i$ are fit
to the e-p data. The $m_i$'s {\em a priori} could also be fit; in
practice, the parameterization has enough flexibility if a fixed set of $m_i$'s
covering  a suitable range is chosen. 
This MD parameterization turns out to be very efficient in fitting the
e-p data. 

The  main interest of the MD parameterization is that an important physics
constraint, explicitly addressed in the VDM, can  be incorporated: requiring
$m_i^2\ge$4$m_\pi^2$ ensures that the individual terms contributing to $G(q)$
fall no  slower in $r$-space than allowed for by the one-pion tail of the most
extended  Fock state of the proton, the  $n$+$\pi^+$
configuration\cite{Hammer04}.  This  constraint ensures from the very beginning
that the large-$r$ density falls in way controlled by physics. 

One might think that the large-$r$ fall off is also affected by the dipole
factor. However, for  $M$ significantly larger than $m_i$, say by a factor of
5, this is not the case. In $r$-space the multiplicative dipole  factor
corresponds to a folding of the (monopole) density with a function that has a width 5 times
smaller than the proton size, typically. While this folding has major effects at
small radii (where it removes the unphysical pole of the monopole-density at
$r$=0), the  folding does not affect the {\em shape} at $r>1fm$.

  We have employed the world data (excluding \cite{Bernauer10b}) up to a
$q_{max}$ of 10$fm^{-1}$, corrected for two-photon effects, using the MD
parameterization. The resulting form factors and radii are very stable upon
variation of those parameters that are not fit. For $q_{max}$=10$fm^{-1}$ and
I=7  we obtain a $\chi^2$ of 544 for 604 data points. The resulting charge-rms
radius amounts to 0.891$\pm$0.013$fm$. For the resulting density see Figure
\ref{rovdma} below.

\subsection{Laguerre polynomial fits \label{lag}}

When attempting to reproduce data over a large range of $q$, it often is not
straightforward to find for the  parameterization of choice the optimal set of
parameter values. The fit may ''get stuck'' in local minima of $\chi^2$ (for a recent
discussion with examples of failed fits see \cite{Bernauer16a}). Moreover, the
parameters could be strongly correlated, a problem which slows convergence. It
therefore might be advantageous to expand the form factor/density in an
orthogonal set of basis functions. 

Among the bases that have been used for heavier nuclei --- Fourier-Bessel,
Hermite functions, Laguerre functions \cite{Friar73,Anni95,Kelly02} --- the
latter is particularly appropriate since it incorporates the exponential-type
fall-off at large $r$ which one expects from the one-pion tail, leading to
densities approximately proportional to $e^{-\mu r}/r$. This physics constraint
avoids, to some degree at least, the general problem with  expansions of $G(q)$
in terms of a complete set of basis functions, namely the fact  that with
contributions at very large $r$  one can fit fluctuations of the data, with
minor reduction in $\chi^2$. 

We accordingly have parameterized the density as 
\bea 
 \rho (r) = \sum_{n=0}^N  a_n ~e^{-x}~ L_n(x) = \sum_{n=0}^{N} a_n 
\sum_{m=0}^n c_{nm}~ x^m~ e^{-x}  
\eea
where $L_n$ is the n$^{th}$ Laguerre polynomial and $x$=r/$\beta$.  The  
corresponding form factor  can  be calculated in closed form, which makes it
easy to fit the  parameters $a_n$  to the data. 
The moments $\langle r^{2n} \rangle$ are given by the first 2$n$+3
coefficients, {\em i.e.} for low $n$ they do not depend on  the order 
$N$ of the polynomial.

As for other multi-parameter expansions, the error bars resulting from the 
Laguerre fits are sensitive to the number $N$ of terms: too many terms allow
for strong correlations between parameters of the highest order driven by the
 fluctuations of the data. This can be avoided, as described in
Section \ref{map},  with a penalty in the $\chi^2$.   

We have made a number of fits of the world data ($q_{max}$=10$fm^{-1}$),
corrected for 2-photon effects, using the Laguerre function \cite{Sick17}. For
2$\cdot$7+1 parameters and 604 data points we find a $\chi^2$ of 540 and an
rms-radius of  0.879$\pm$0.02$fm$, with a density which agrees out to $2.6fm$
with the ones shown in Figure \ref{rovdma}.

\subsection{Sum-Of-Gaussians with tail-constraint \label{sog}}

For nuclei $A \ge 2$ the SOG parameterization \cite{Sick74}  has  been employed
often. Here, the density is parameterized as a  sum of (symmetrized) gaussians
placed at many different radii. The width of the gaussians limits the fine
structure of  $\rho (r)$ (relevant for $\delta \rho$, but not for $R$). As
compared to other parameterization of $\rho (r)$ or $G(q)$ SOG features the
strongest decoupling of different radii. In order to get a realistic behavior of
the density at large radii --- the importance of which has been stressed
repeatedly above --- we have used SOG for the proton by including a tail
constraint.

At small radii, say $r<1fm$, the quark/gluon structure of the proton is very
complicated. At {\em large} $r$, one expects the density to be dominated by the
Fock component with the smallest separation energy, the $\pi^+$+neutron
configuration. As an example, we cite the cloudy bag model where the density
outside the bag radius of $r \sim 0.8 fm$ is entirely given by this 1-$\pi$
tail.

\begin{figure}[htb]
\centering
\includegraphics[scale=0.4,clip]{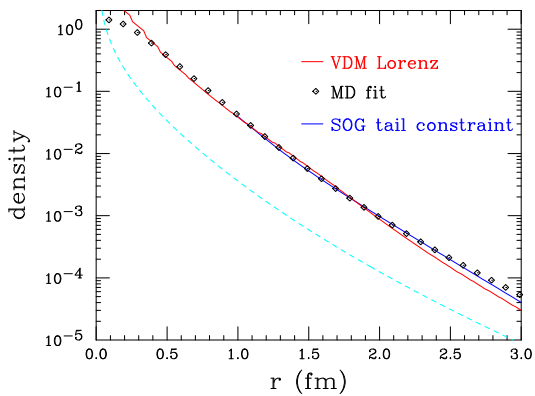}  
\caption{\label{rovdma} The experimental charge density, obtained from the 
MD fit ($\diamond$) of the world data is compared to the tail of the SOG fit
(blue) 
\cite{Sick12} and the density corresponding to the VDM fit of Lorenz {\em et
al.} (red). Not shown for clarity: the density from \cite{Graczyk14}, see
Section \ref{gra}.}
 \end{figure}

The {\em shape} of the corresponding density --- but not the absolute
normalization --- can be calculated from the asymptotic radial wave function
$W_{-\eta, 3/2} (2 \kappa r)/r$ of the pion, with $\kappa$ given by the pion
mass and removal energy. This shape can be used as a constraint on the shape of
$\rho (r)$ fit to the e-p data. In \cite{Sick12} details and corrections to this
simple prescription are discussed (they are of minor numerical importance). The
physics of this constraint is the same as addressed by the $2\pi$
triangle-diagram in the VDM. For examples of similar tail constraints in
analyses of A$\ge$2  data --- which yield the most precise radii which all agree
with the muonic X-ray results --- see \cite{Sick98,Sick08,Sick82}.

In Figure \ref{rovdma} we compare various densities: the Fourier transform of
$G_e(q)$ ({\em i.e.} the density without relativistic corrections) of the MD fit,  the
density corresponding to the VDM fit of Lorenz {\em et al.}\footnote{The tail 
density is {\em much} higher than the 1-$\pi$ tail deduced 
initially
in \cite{Hammer04} (dashed)}\cite{Lorenz12b,Lorenz14} (which includes the relativistic correction discussed in
Section \ref{randro} in order to make it comparable) and the tail density of the
SOG fit. While the large-$r$
shape of the MD density and the density calculated from the $n\pi^+$ Fock
state are very close, the
tail the VDM density falls somewhat more quickly, as a consequence of the low
near-threshold strength of the spectral function.

A fit of the world data ($q_{max}$=10$fm^{-1}$) including 2-photon corrections
using the SOG density together with the tail constraint for $r>1.2fm$ yields
\cite{Sick12} a charge  rms-radius of $R$=0.886$\pm$0.008$fm$. The tail
constraint is shown in Figure \ref{rovdma}.        

\section{Summary}

In this paper we have discussed  approaches used by various authors to extract
the proton charge rms-radius $R$ from data on elastic electron-proton
scattering.  We have pointed out the difficulties of the standard approach of
fixing the attention exclusively  upon the low-$q$ data and   the slope of the
electric form factor $G_e(q)$ at $q$=0. The crux lies in the extrapolation from
$q$'s, where the data are sensitive to $R$, to $q$=0. The curvature of the
parameterized $G_e(q)$'s needed for this extrapolation introduces a model
dependence; the parametrizations in particular lack the physics constraint that
the form factor is (with or without relativistic corrections) the Fourier
transform of a density confined to a  region in $r$ of, say, $r<3fm$.

We have emphasized that this curvature at low $q$ is related to the shape of the
charge density $\rho (r)$ at large radii $r$. About the latter quantity we do
have knowledge from physics, contrary to the  curvature of $G(q)$ at low $q$;
the density at {\em large} $r$ is dominated by the least-bound Fock state of the
proton, the $\pi^+n$ configuration. To obtain a reliable $R$ one must make sure
that the density corresponding to $G(q)$  is reasonably close to this behavior.

This entails three consequences:

--- Use of a parameterized $G_e(q)$ that is {\em physical}, {\em i.e.} does
indeed correspond to a density. This is not the case for most parameterizations
employed in the literature. 

--- Fit of the data to the largest $q_{max}$, in which case the data themselves 
fix to a fair degree the shape of $\rho (r)$  {\em including} its behavior at
large $r$, hereby constraining the shape of $G_e(q)$ at low $q$.

--- Verification that $\rho (r)$ at large $r$ shows a physical behavior and,
better, use of a physical constraint to enforce the correct behavior. The fall-off
given by the pion tail provides a very general and helpful {\em physics}
constraint. 

We have discussed a number of analyses of the e-p data that do respect the above
insights. To determine the radius we use the unweighted average of the (in part
redone) fits: [3/5]Pad\'e \cite{Arrington07}, dipole times polynomial in $z$
\cite{Lee15}, dipole times polynomial in $\xi$ \cite{Borisyuk10}, Bayesian
inference \cite{Graczyk14}, MD, Laguerre  and SOG \cite{Sick12}. 
This yields a radius of
\bea
 R = 0.887 \pm 0.012 fm, 
\eea 
where the error  bar covers  both
the uncertainties of individual results as well as  their scatter. 
The results from the various analyses of e-p scattering which use {\em physical}
parameterizations that lead to a sensible density at large $r$ are quite
compatible. 

\acknowledgments{I want to thank Hans-Otto Meyer and Dirk Trautmann for many
helpful discussions.}


\end{document}